\documentclass[twocolumn]{aastex6}

\newcommand{\figscaleone}{\epsscale{1.15}} 

\received{22nd March 2017}

\shorttitle{Scattering associated with stars}
\shortauthors{Walker et al}

\begin{document}

\title{Extreme radio-wave scattering associated with hot stars}
\author{Mark A. Walker$^*$}
\affil{Manly Astrophysics, 15/41-42 East Esplanade, Manly, NSW 2095, Australia}
\email{$*$ Mark.Walker@manlyastrophysics.org}
\author{Artem V. Tuntsov}
\affil{Manly Astrophysics, 15/41-42 East Esplanade, Manly, NSW 2095, Australia}
\author{Hayley Bignall$^\dagger$}
\affil{CSIRO Astronomy and Space Science, 26 Dick Perry Avenue, Kensington, WA 6151, Australia}
\email{$\dagger$ hbignall@gmail.com}
\author{Cormac Reynolds}
\affil{CSIRO Astronomy and Space Science, 26 Dick Perry Avenue, Kensington, WA 6151, Australia}
\author{Keith W. Bannister$^\ddagger$}
\affil{CSIRO Astronomy and Space Science, PO Box 76, Epping NSW 1710, Australia}
\email{$\ddagger$ Keith.Bannister@csiro.au}
\author{Simon Johnston}
\affil{CSIRO Astronomy and Space Science, PO Box 76, Epping NSW 1710, Australia}
\author{Jamie Stevens}
\affil{CSIRO Paul Wild Observatory, 1828 Yarrie Lake Road, Narrabri, NSW 2390, Australia}
\author{Vikram Ravi}
\affil{Cahill Center for Astronomy and Astrophysics, MC 249-17, California Institute of Technology, Pasadena, CA 91125, USA}

\begin{abstract}
We use data on extreme radio scintillation to demonstrate that this phenomenon is associated with hot stars in the solar neighbourhood. The ionized gas responsible for the scattering is found at distances up to $1.75\,{\rm pc}$ from the host star, and on average must comprise $\sim10^5$ distinct structures per star.  We detect azimuthal velocities of the plasma, relative to the host star, up to $9.7\,{\rm km\,s^{-1}}$, consistent with warm gas expanding at the sound speed. The circumstellar plasma structures that we infer are similar in several respects to the cometary knots seen in the Helix, and in other planetary nebulae. There the ionized gas appears as a skin around tiny molecular clumps. Our analysis suggests that molecular clumps are ubiquitous circumstellar features, unrelated to the evolutionary state of the star. The total mass in such clumps is comparable to the stellar mass.
\end{abstract}


\section{Introduction}
The refractive index of ionized gas is very large at low frequencies, leading to profound effects on the propagation of radio-waves through interstellar space --- see the review by \citet{rickett1990}. A useful reference model for many of the observed effects is distributed Kolmogorov turbulence in the diffuse, ionized interstellar medium (ISM) \citep[e.g.][]{armstrong1995}. However, there are several phenomena, which collectively can be called ``extreme scattering'', that are known to involve flux modulations and scattering angles much larger than can be accommodated in the Kolmogorov picture. Manifestations of extreme scattering include: multiple imaging of pulsars \citep{cordeswolszczan1986,rickett1997}; Extreme Scattering Events \citep[ESEs:][]{fiedler1987,fiedler1994,bannister2016}; intra-day variability of flat-spectrum radio quasars \citep[IDVs:][]{kedziora1997, dennettthorpe2000,bignall2003}; and parabolic arcs in pulsar ``secondary'' spectra \citep{stinebring2001,cordes2006}. These phenomena share at least one key characteristic: they all appear to require plasma pressures that are much greater than is typical of the diffuse ISM, and it therefore seems likely that they are somehow related. At present there is no consensus on the cause of any of these phenomena, with suggestions ranging from current sheets seen edge-on \citep{goldreichsridhar1995,penlevin2014}, to the ionized skins of tiny, molecular clouds \citep{walkerwardle1998,walker2007}.

Part of the difficulty in studying extreme scattering is that the phenomenon is transient, so one must first identify an active source, and then bring telescopes to bear on it. With a new observing program, dubbed ATESE, on the Australia Telescope Compact Array (ATCA), we have been able to identify sources undergoing extreme scattering, in a way that makes efficient use of telescope resources \citep{bannister2016}. Although the ATESE project aims to find ESEs in progress, and then study them intensively, we have also identified new IDV sources \citep[e.g.][]{tuntsov2017}.

Most recently, the ATESE project discovered  strong variability in the source PKS1322-110 \citep{bannister2017}. Subsequent follow-up with ATCA showed large-amplitude oscillations on a timescale of a few hours, so we now classify it as a (rapid, large-amplitude) IDV source. While considering optical follow-up of PKS1322-110, we realised that it lies only $8^\prime\!.5$ from Spica ($\alpha$ Vir) --- the 16th brightest star in the sky. Only a handful of extreme scintillators are known, so it seemed improbable that this close positional coincidence would occur by chance. We were also aware of the suggestion of \citet{dennettthorpe2002}, that the IDV of J1819+3845 might be related to the foreground, bright star Vega. We therefore examined whether ionized gas associated with foreground stars provides a good explanation for the IDV phenomenon.

 Although we continue to monitor PKS1322-110, it will be some time before that data set rivals the coverage of published data on the two best studied IDV sources: J1819+3845 \citep{dennettthorpe2000,dennettthorpe2002,dennettthorpe2003}, and PKS1257-326 \citep{bignall2003, bignall2006}. Both of these sources displayed IDV over an interval of many years, revealing annual cycles in the scintillation timescale (which follow from the changing velocity of the Earth's orbit around the Sun). The extensive, high quality data for both sources provide tight constraints on the plasma responsible for the scattering, permitting stringent tests of any model that might be proposed, and they form the basis for our analysis.

This paper is organised as follows. In the next section we present a rationale for selecting candidate stellar counterparts to extreme scintillators, and apply it to the two well studied IDV sources J1819+3845 and PKS1257-326, yielding Vega and Alhakim, respectively, as likely stellar counterparts. The positions and motions of these stars suffice to predict the annual cycles of IDV, and we find that those predictions are a good match to the observed annual cycles. In \S3 we show that it is highly improbable that this agreement is fortuitous. Section 4 considers the physical context in which the plasma structures might arise, and identifies the cometary knots in the Helix Nebula as the appropriate paradigm.  Discussion and conclusions follow in \S\S5,6.

\section{Hot stars foreground to IDV sources}
The discovery of IDV in PKS1322-110 \citep{bannister2017}, with Spica less than $9^\prime$ away in the foreground, immediately suggests an association between the star and the ionized gas responsible for the scattering. The fact that Spica is a (binary) B-star further suggests that hot stars may be necessary to create the conditions for extreme scattering, because B-type (and earlier) stars constitute a tiny fraction of all stars. Reinforcing that point is the idea that UV photons might be the agent that produces the plasma. We therefore adopt the hypothesis that material associated with hot, foreground stars is responsible for extreme scattering. The single example of Spica does not tell us which spectral types may give rise to extreme scattering. Initially we include types O, B and A in our definition of ``hot'' stars; we return to this issue later in the paper (\S5).

To be more specific: our hypothesis is that IDV is caused by radially-elongated plasma structures around hot stars. The rationale for specifying a radial orientation will become clear later. That the structures responsible for IDV must be elongated (i.e. statistically anisotropic) was pointed out by Barney Rickett (personal communication to MAW) soon after IDV was discovered in PKS0405-385  \citep{kedziora1997,rickett2002}. It was subsequently confirmed for J1819+3845 and PKS1257-326.\footnote{It is also true of pulsar parabolic arcs -- see \citet{walker2004,cordes2006,brisken2010} -- which appear to be the counterpart, in pulsars, of the IDV phenomenon in quasars.} Indeed, the annual cycles of both of these sources are well approximated by a model in which the anisotropy is infinite --- i.e. a one-dimensional model \citep{walker2009}. We employ the one dimensional approximation in this paper.

In order to identify which foreground stars could be responsible for the scintillation of a background radio source, we need to define the domain of influence of each star. Various physically-motivated definitions are possible, in principle --- e.g. based on the ionizing luminosity of the star. Our choice is the Hill surface, which defines the region within which material can remain bound to the star, in the presence of the Galactic tidal field. For the Sun, the Hill surface has principal semi-axes of approximately $(1.42, 0.92, 0.74)\,{\rm pc,}$ in cylindrical, Galactic coordinates $(R, \phi, Z)$ \citep{antonovlatyshev1972}. The Galactic tidal field for local stars is similar to that experienced by the Sun, but the volume of the Hill surface scales with the stellar mass. In this paper we therefore adopt a fiducial impact parameter of $2\,{\rm pc}$, as the domain of influence of local, massive stars.

\begin{figure}
\figscaleone
\centerline{\plotone{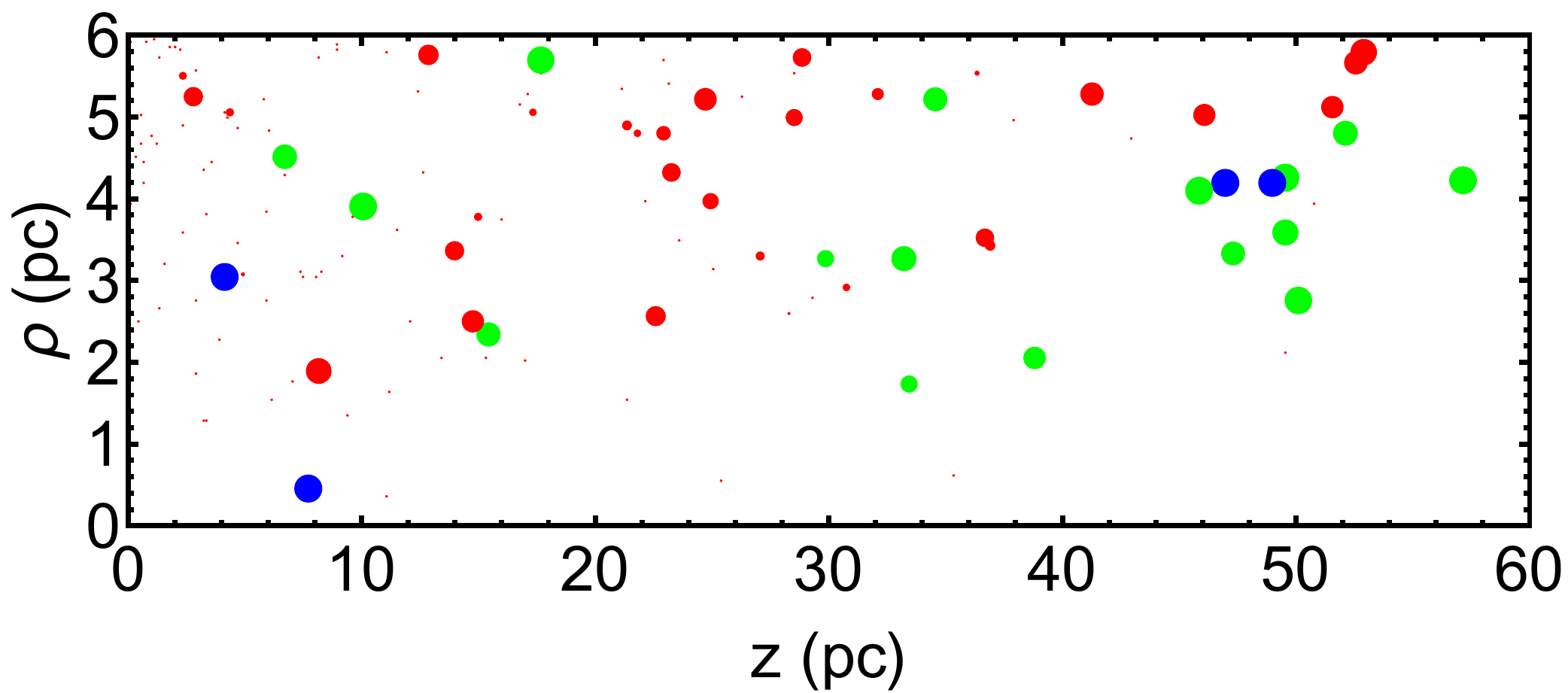}}
\centerline{\plotone{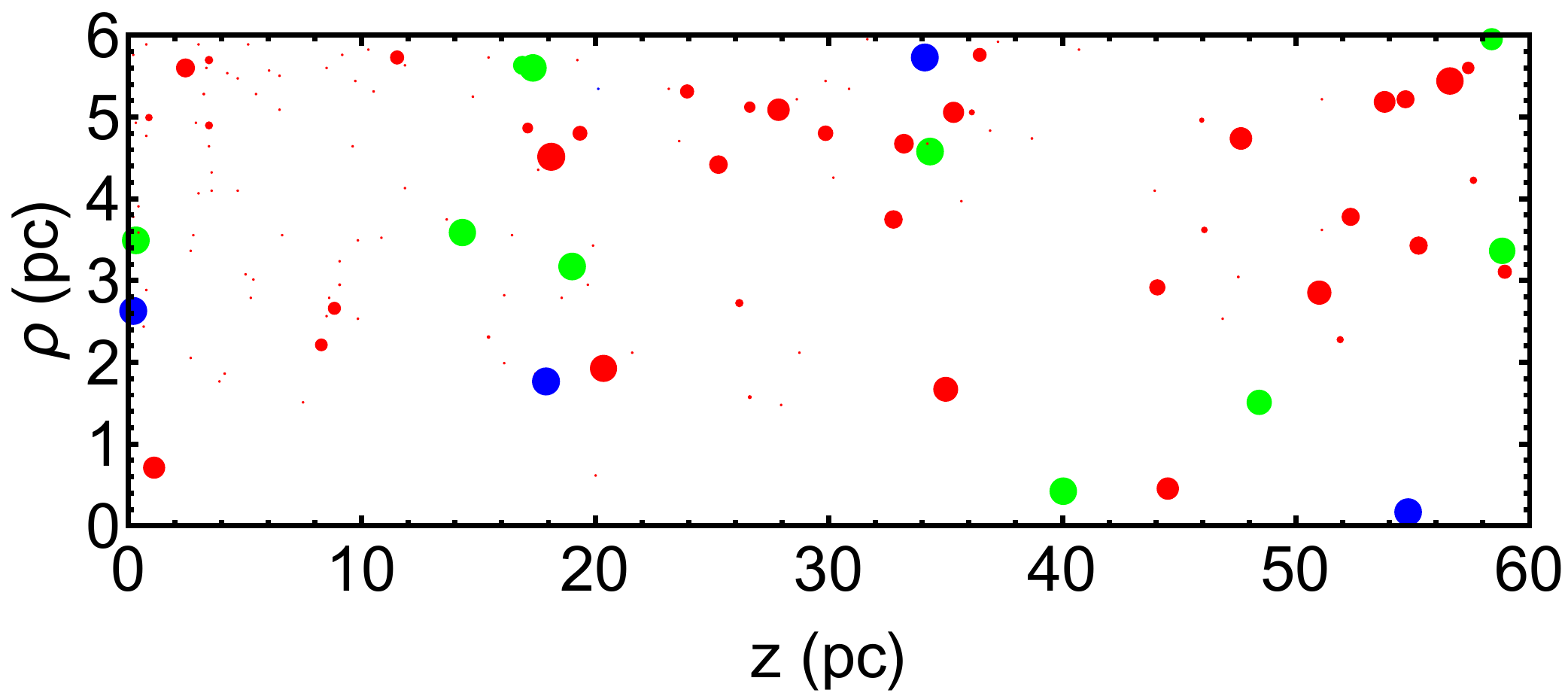}}
\caption{The locations of Hipparcos stars, in cylindrical coordinates, $(\rho,z)$, around the line-of-sight to J1819+3845 (top), and PKS1257-326 (lower panel). The size of each point represents luminosity, and colour indicates spectral type: blue $\leftarrow$ (O, B, A);  green $\leftarrow$ (F);  red $\leftarrow$ (G, K, M).}
\end{figure}

\begin{table*}
\caption{Properties of scattering plasma, and of nearby stars.}
\centering
\begin{tabular}{lcccc|ccccr}
\tableline
                          &  $\beta$ &  $v_\perp$  &  $a_\perp$ &  $z_s$ & $\beta^*$ & $v_\perp^*$ & $z^*$ & $\rho^*$ &  \cr
Radio Source    &$\!\!\!\!$ ${\rm (N\rightarrow E)^\circ}$ &$\!\!\!\!$ ${\rm(km\,s^{-1})}$ & ${\rm(Mm)}$ &  ${\rm(pc)}$  & ${\rm (N\rightarrow E)^\circ}$ & ${\rm(km\,s^{-1})}$ & ${\rm(pc)}$ & ${\rm(pc)}$ &$\!\!\!\!$ Star\cr
\tableline
J1819+3845      &$\!\!\!\!$ $-97.35\;(0.4)$  &$\!\!\!\!$ $19.67\;(0.24)$  &$\!\!\!\!$ $29.5\;(0.6)$ & $10.9$   & $-91.93\;(0.00)$ &$\!\!\!\!$ $10.02\;(0.07)$ &$\!\!\!\!$ $7.74\;(0.03)$ &$\!\!\!\!$ $0.461\;(0.002)$ &$\!\!\!\!$ Vega\cr
PKS1257-326    &$\!\!\!\!$  $126.60\;(0.6)$ &$\!\!\!\!$ $ 20.50\;(0.30)$ &$\!\!\!\!$ $44.9\;(1.0)$ & $13.5$    & $134.50\;(0.00)$ &$\!\!\!\!$ $ 22.4\;(0.4)$ &$\!\!\!\!$ $17.9\;(0.2)$ &$\!\!\!\!$ $1.75\;(0.02)$ &$\!\!\!\!$ Alhakim\cr
\tableline
\end{tabular}
\tablecomments{Columns 2, 3 and 4 are from Table 1 of \citet{walker2009}; they characterise the scattering plasma. Column 5 gives $z_s=a_\perp/\theta_\perp$, with $\theta_\perp$ as specified in \S2. Columns 6, 7, 8 and 9 are the azimuth, perpendicular velocity component, and distances along and perpendicular to the line-of-sight for the host stars. Measurement uncertainties are given in parentheses; the uncertainty associated with $z_s$ is predominantly systematic, and is likely $\sim0.25\,{\rm dex}$.}
\end{table*}

The very rapid scintillations of J1819+3845 and PKS1257-326 leave no doubt that the scattering material must be located at line-of-sight distances $\ll1\,{\rm kpc}$ from us \citep{dennettthorpe2000,dennettthorpe2003}. Thus a deep stellar catalogue is not necessary for cross-matching; we employ the Hipparcos catalogue \citep{perryman1997}. Figure 1 shows the local Hipparcos stars along the line-of-sight to our two IDV sources, in cylindrical coordinates $(\rho,z)$ relative to the axis defined by the direction to the radio source, and with the observer at the origin. We see that the line-of-sight to J1819+3845 pierces the Hill surface of only one, local, hot star -- Vega ($\alpha$ Lyr, ${\rm A0V},\,z=7.7\,{\rm pc}$) -- whereas for PKS1257-326 there are two: Alhakim ($\iota$ Cen, ${\rm A2V},\,z=18\,{\rm pc}$), and HD112934 (${\rm A9V},\,z=55\,{\rm pc}$). That these lines of sight pass through the Hill surfaces of three hot stars with $z<60\,{\rm pc}$ is, in itself, slightly surprising. The cylindrical regions shown in Figure 1 encompass a volume $9\times$ greater than the volume of interest, and together they contain 8 hot stars (blue points, in the figure). We would therefore expect to see $0.9$ hot stars within $\rho<2\,{\rm pc}$ from these lines-of-sight, whereas we find three.

Although three hot stars meet our selection criterion, for the present we set aside HD112934 and concentrate on Alhakim and Vega. The specific reasons for doing so will become clear in \S5, where we argue that HD112934 is probably the chance interloper we expected to find.

The coincidences between our radio sources and their foreground stars become much more remarkable when we consider all the constraints on the scattering plasma, coming from the annual cycles of the IDV. These constraints localise the scattering material in three spatial dimensions, one angle and one velocity component, and are listed in Table 1. Corresponding quantities for the proposed counterpart stars are also given, as per the Hipparcos catalogue. The quantities listed are as follows. The angle $\beta$ is the orientation of the long axis of the (highly anisotropic) flux pattern; whereas $\beta^*$ is the orientation of the line joining the star and the radio source. The quantity $v_\perp$ is the (negative of) the velocity component of the scattering plasma, perpendicular to the long axis of the flux pattern; whereas $v_\perp^*$ is the (negative of) the velocity component of the star, perpendicular to the same axis.\footnote{The velocity, $v_\perp$, as defined by \citet{walker2009}  is the steady component of the observer's velocity through the flux pattern, and in the context of the present discussion that corresponds to the negative of the relevant component of the plasma velocity, relative to the solar system barycentre.} The quantity $a_\perp$ is the length-scale of the flux pattern perpendicular to the long axis, determined from the product of the instantaneous perpendicular speed and the timescale of the scintillations.

For a point-like radio source, the spatial flux pattern created by the scattering material would have structure on a wide range of scales, with large fluctuations on short timescales, for example. For a quasar, the source-size has a smoothing effect on the observed flux pattern: all fluctuations on angular scales small compared to the source size are heavily damped, and $a_\perp=z_s\theta_\perp$, where $\theta_\perp$ is the angular size of the  autocorrelation of the source intensity profile, and $z_s$ is the distance of the scattering material from the observer. To evaluate $\theta_\perp$ we model the radio emission as a synchrotron source near the self-Compton limit, mildly boosted towards the observer: a circularly-symmetric Gaussian intensity profile, $\exp(-\theta^2/2\sigma^2)$, with a peak brightness temperature of $10^{13}\,{\rm K}$. Both radio sources have mean fluxes at $5\,{\rm GHz}$, where the scintillation timescales were measured, of approximately $200\,{\rm mJy}$, from which we determine $\sigma\simeq13\,{\rm\mu as}$. The uncertainty in this estimate is likely $\sim0.25\,{\rm dex}$, associated with $\sim0.5\,{\rm dex}$ uncertainty in the peak brightness temperature.

We require $\theta_\perp$ appropriate to the reported scintillation timescales, which follow different conventions depending on the author (e.g. the half-width at half-maximum of the temporal autocorrelation of the flux measurements). The models reported by \citet{walker2009} were fit to timescales corresponding to $\theta_\perp\simeq1.35\sigma$, for J1819+3845 \citep[Table 1, column 8]{dennettthorpe2003}, and $\theta_\perp\simeq1.67\sigma$, for PKS1257-326 \citep{bignall2003}. This model yields the estimate $z_s=a_\perp/\theta_\perp$, as given in column 5, Table 1. The corresponding line-of-sight distance of the stars is given in column 8. 

Using Table 1, we can compare the values of $\beta$, $v_\perp$ and $z_s$, describing the scattering plasma, with those of $\beta^*$, $v_\perp^*$ and $z^*$, describing the positions and motions of the proposed counterpart stars. Three things are immediately evident from that comparison. First, the differences in orientations and velocity components are large compared to the uncertainties in the measurements. In \S4 we discuss the interpretation of those differences. Although values for $a_\perp$ are determined quite precisely from the scintillation data, the same cannot be said of $z_s$, because of the systematic uncertainty associated with our simplistic model of the radio source structure. Secondly, although our source model is unsophisticated, it suffices to demonstrate that the scattering material must be very local, as our counterpart stars indeed are. Thirdly, the orientations $\beta$ are similar to the azimuths of the stars, $\beta^*$. It's less obvious whether the few ${\rm km\,s^{-1}}$ differences between $v_\perp$ and $v_\perp^*$ are surprising. In the next section we address that issue, and we assess the probability that Vega and Alhakim are not responsible for the observed scintillations; it is very small.

\section{Probability of chance associations}
Consider the possibility that the hot stars we have identified, close to the line-of-sight to IDV sources, have no causal relationship to the scintillations. In that case, we are drawing randomly from amongst all such local stars, and we can use the Hipparcos catalogue to estimate the chances of two random coincidences that are at least as close as the associations we have identified. We define ``local'', here, to mean stars within $60\,{\rm pc}$.

We assume that the position and velocity distributions are approximately independent, locally. We characterise the former by the local density of hot stars, $n=4.0\times10^{-4}\,{\rm pc^{-3}}$. For the latter, we require the probability of matching specific velocity components, perpendicular to the line-of-sight. As the local, hot stars are found all over the sky, we take the relevant transverse velocity component to be in the same orientation, with respect to Galactic coordinates, as for the scintillation patterns.\footnote{In principle, a better estimate could be arrived at by using the three-dimensional vector component, parallel to that measured for the scintillation pattern.  However, reliable radial velocity information is often not available for hot stars.} From Table 1, the velocity intervals of interest are (in ${\rm km\,s^{-1}}$) $v_\perp=19.7\pm9.7$, for J1819+3845, and $v_\perp=20.5\pm1.9$, for PKS1257-326, within which there are 85 and 21 local, hot stars in the Hipparcos catalogue, respectively, out of a total of 365. Thus we find the probability of a chance match in velocity, at least as close as we obtain for J1819+3845-Vega, to be $p_\perp=0.233$; and for PKS1257-326-Alhakim it is $p_\perp=0.058$.

The number of stars that we expect to match the scattering plasma, in position and velocity, at least as well as our chosen stars, is just (with $\Delta\beta$ in radians)
\begin{equation}
N=n\,\Delta\beta\,\rho^2\,\Delta z\,p_\perp.
\end{equation}
The relevant intervals $\rho\le\rho^*$, $\Delta\beta=2|\beta^*-\beta|$, and $\Delta z=2|z^*-z_s|$  can be determined from Table 1. We thus find that $N=2.4\times10^{-5}$ for the J1819+3845-Vega pairing, and $N=1.7\times10^{-4}$ for PKS1257-326-Alhakim. It is, therefore, unlikely that either of our proposed associations would have arisen as a random coincidence.

Taken together, the two proposed pairings are extremely unlikely to have arisen by chance. To evaluate the overall probability, we must consider how our original hypothesis -- i.e. that the scattering material is radially oriented filaments, associated with hot stars -- was arrived at. It was the link between Spica and PKS1322-110 that triggered our investigation of a connection with hot stars; that piece of information is independent of the probabilities we have computed here, and thus has no impact upon them. However, the notion that the anisotropic structure of the scattering material is radially organised did not come from the PKS1322-110-Spica connection. Rather it was suggested to us by the fact that $\beta$, for J1819+3845, is similar to $\beta^*$ for Vega. Thus the overall probability of obtaining our result by chance is increased by the ratio $\pi/\Delta\beta$ for the J1819+3845-Vega association --- a factor of 16.6. In conclusion, the probability of obtaining both results by chance is $6.8\times10^{-8}$.

Having established that there is a physical association between the foreground stars and the IDV, we now consider the origin of the plasma structures.

\section{Nature of the plasma filaments}
Perhaps the first physical picture that comes to mind is that the filamentary structures might arise in a wind from the central star. There are several problems with that idea. First, A-star winds are feeble. An upper limit on the  mass loss from Vega, for example, is $\dot{M}\la4\times10^{-10}\,{\rm M_\odot\,yr^{-1}}$ \citep{lanzcatala1992}, which corresponds to mean electron densities $n_e\la10^{-6}\,{\rm cm^{-3}}$ at $2\,{\rm pc}$, if the wind velocity is $300\,{\rm km\,s^{-1}}$. By contrast, the IDV phenomenon requires values $\sim10\,{\rm cm^{-3}}$ \citep{tuntsov2013}; the stellar wind is not a natural source of such dense plasma. We would thus require the wind to be concentrated into radial streamers, covering a miniscule fraction of the sphere, and with a density contrast $\ga10^7$. If the wind were a purely hydrodynamic phenomenon, such density contrasts should dissipate rapidly in a turbulent cascade. It is tempting to invoke magnetic fields, in a hydromagnetic wind, strong enough to confine the dense plasma. If so, the field lines and the plasma filaments ought to be azimuthally oriented at large radii \citep{parker1958}, whereas we require radial filaments. In short, it seems unlikely that the hot star is the source of the scattering plasma.

Having excluded the stellar wind, we are led to the idea that the source of the plasma is itself remote from the star, but close to the line-of-sight to the background radio source. The central stars are, however, a source of UV photons in abundance, so one possibility is that the plasma could be an ionized flow driven from the surface of dense, neutral clumps, bound to the star. Such clumps have no counterpart in the theory of stellar structure and evolution. There is, however, at least one observed example of a strikingly similar situation: the cometary knots in the Helix Nebula \citep[e.g.][]{meaburn1992,odellhandron1996,huggins2002,meixner2005,matsuura2009}. Such structures appear to exist in other planetary nebulae \citep{odell2002}, but are most clearly seen in the Helix, presumably because it is one of the closest planetaries.

Several points of similarity argue that the cometary knots in the Helix provide the correct paradigm for the environments of hot, main sequence stars. First, the cometary knots are seen at distances of $0.2$-$0.4\,{\rm pc}$ from the central star of the nebula -- see, particularly, \citet{matsuura2009} -- this is not much smaller than is required for the structures around Vega ($0.45\,{\rm pc}$). And it is possible that the Helix might contain yet more knots at greater distances from the central star. Being shielded from the stellar radiation by gas at smaller radii, they would be harder to see, and could fall below the detection limit of observations to date. Secondly, extreme scattering is a phenomenon that is typically transient \citep{kedziora2006,lovell2008,debruynmacquart2015}, and is thus likely to be caused by discrete entities; the cometary knots in the Helix, although large in number, appear to be genuinely discrete. Thirdly, the transverse scale of the cometary knots in the Helix is $\sim10^2\,{\rm AU}$; this is very tiny by astronomical standards, but similar to the upper end of the range of dimensions inferred for extreme scattering structures. In particular, the plasma concentration responsible for IDV in J1819+3845 is $\sim10^2\,{\rm AU}$ in size \citep{debruynmacquart2015}. Fourthly, the tails of the cometary knots in the Helix are everywhere approximately radial, matching the orientation required for the IDV plasma anisotropies. Although the observed morphology of individual cometary knots corresponds to spatial scales that are $\ga10^5\times$ greater than the plasma inhomogeneities responsible for IDV (i.e. $a_\perp$, in Table 1), the anisotropy on small scales may well be determined by the large-scale morphology and/or flow patterns of the plasma. Finally, the incidence of rapid, large amplitude IDV requires $\sim10^5$ plasma concentrations around each hot star --- consistent with the number of cometary knots identified in the Helix \citep{matsuura2009}. To see this we calculate the probability of a randomly chosen line-of-sight intersecting one of the plasma concentrations responsible for IDV. The size of the latter has been estimated at $\sim100\,{\rm AU}$ for J1819+3845 \citep{debruynmacquart2015}, so we adopt a cross-section $\sim10^4\,{\rm AU^2}$. With a hot-star number density of $4\times10^{-4}\,{\rm pc^{-3}}$, and a path length (scale height of hot stars in the Galactic disk) $\sim100\,{\rm pc}$, we would have a sky-covering fraction of $\sim10^{-8}$ if each hot star carried a single plasma concentration. The actual sky-covering fraction can be estimated from the results of \citet{lovell2003}, who found that J1819+3845 was the only rapid, large-amplitude variable in a sample of 710 compact radio quasars; hence $\sim10^5$ plasma concentrations are required around every hot star.

Returning now to Table 1, the paradigm provided by the Helix facilitates straightforward interpretations of the differences $|\beta-\beta^*|$, and $|v_\perp-v_\perp^*|$, both of which are several times larger than their  measurement uncertainties. The observed differences in orientation can arise in two ways: the symmetry axes of the cometary tails may deviate from precisely radial orientations, and the individual tails may have a non-zero opening angle. We note that foreshortening tends to increase the apparent deviation -- e.g. by a factor of 2 at the median inclination of $60^\circ$ -- thus the intrinsic deviation from radial is probably $\la4^\circ$, in both the symmetry axes and the tail opening angles. The observed differences in perpendicular velocity cannot be due to the azimuthal velocities of the neutral clumps, relative to the stars, as they could not then be bound to their hosts. Instead, we should ascribe the detected azimuthal velocity components to motion of the plasma relative to the clumps. The relatively large value of $9.7\,{\rm km\,s^{-1}}$, inferred for J1819+3845-Vega, is suggestive of free expansion of the plasma in the azimuthal direction, perpendicular to the cometary tails.

\section{Discussion}
The essential similarities between plasma structures in the Helix nebula and those around nearby stars indicate that neutral gas clumps, which are known to be molecular in the case of the Helix \citep{huggins1992}, are a common circumstellar feature. Although we drew that connection in order to illuminate the origin of IDV, it also has implications for the Helix. If swarms of molecular clumps are present around main sequence stars, they did not get there as a result of late-stage stellar evolution, as is often assumed for the Helix \citep[e.g.][]{garciasegura2006}. We will argue that main sequence stars of all types may carry populations of molecular clumps. The progenitor of the Helix could have been such a star, so it is no longer appropriate to assume that the observed clumps are a product of stellar mass-loss.

Where, then, did all of these clumps come from? Viable formation scenarios must precede the main sequence. One possibility is that clumps form contemporaneously with stars. Another possibility is that parsec-sized clusters of molecular clumps constitute the starting point, and a star subsequently forms at the centre of each cluster as a result of disruptive collisions between clumps. We cannot say which, if either, of these speculations is correct. There are two key considerations for any proposed scenario. First, the clumps must have long lifetimes -- comparable to, or greater than the main-sequence lifetime of an A-star -- which strongly suggests that they are, individually, self-gravitating clouds. Secondly, the formation mechanism must be very efficient, because the total mass in molecular clumps is large --- as we now demonstrate.

In the case of the cometary knots in the Helix, estimates of the masses of individual, neutral clumps have yielded values $\sim2\times10^{-5}\,{\rm M_\odot}$ \citep[e.g.][]{meaburn1998}, implying that their total mass in the Helix is $\ga1\,{\rm M_\odot}$. This estimate is very rough, as both clump masses and numbers are highly uncertain. Notwithstanding those uncertainties, it appears that the total mass is comparable to the stellar mass, and in the present work we have argued that the situation is similar for local, hot stars. As this paper concerns itself with the interpretation of IDV, which involves a strong selection bias in favour of local, ionized material, we anticipate that stars of all spectral types, throughout the Galaxy, may also be surrounded by swarms of tiny molecular clumps. If so, then the total molecular clump mass constitutes a substantial fraction of the mass of the Galaxy.

That conclusion was also reached by Walker and Wardle (1998), via a different line of reasoning. They demonstrated that if ESEs are caused by the ionized skins of self-gravitating, neutral clouds, then the observed event rate requires that those clouds must form a substantial component of the Galaxy's dark matter. Detailed models of such clouds confirm that they can have very low luminosities (Walker and Wardle, in prep.). 

A third line of evidence supports the view that the total mass of molecular clumps is large. Using recent results for the Oort constants, based on {\it Gaia\/} data \citep{bovy2017}, and the local mass density determined from {\it Hipparcos\/} data \citep{creze1998}, we have re-evaluated the size and shape of the Hill surface, following \citet{antonovlatyshev1972} and \citet{heislertremaine1986}. We find principal semi-axes of $(1.37, 0.91, 0.73)\,{\rm pc}$, for the Sun. Taking a mass of $2.5\,{\rm M_\odot}$ for Alhakim \citep[HD115892:][]{quanz2011}, we find that the Hill surface extends only 70\% of the angular distance to PKS1257-326. We therefore require an additional mass $\ga5\,{\rm M_\odot}$, bound to Alhakim, in order to extend the Hill surface as far out as the PKS1257-326 line-of-sight. That conclusion is specific to Alhakim, but, in combination with our other two estimates of molecular mass, it alludes to a likely conflict with contemporary cosmological models, in which most of the matter is non-baryonic.

How do our results sit in relation to ideas about the local ISM? It has been suggested that the IDV of J1819+3845 and PKS1257-326 is connected with the boundaries of certain local clouds, whose properties have been inferred from absorption-line spectroscopy \citep{linsky2008}. The absorption-line clouds have much larger sky-covering fractions than the plasma concentrations responsible for extreme scattering, and are therefore not directly identifiable with the cometary knots themselves. However, as the plasma continues to flow outward, away from the star, it presumably expands laterally until it becomes space filling. The resulting gas would have a low density, but the column-density might nevertheless be high enough to be detectable in absorption-line studies. It might therefore be interesting to re-examine the absorption-line data to see whether they are consistent with, e.g., centro-symmetric velocity fields around local stars.

It is interesting to consider how the scintillations of background radio sources might change when the molecular cloud clusters are located around late-type stars, rather than the early types we have been concerned with so far. Lower photospheric temperatures combined with lower bolometric luminosities (for main sequence stars) mean that the UV radiation fields are much smaller, for the late-types. Likewise, the late-types are expected to have lower mechanical (wind) luminosities, at least while stars are on the main sequence. Consequently the plasma skins are expected to be much less prominent in the case of molecular clumps around cooler stars. Presumably this means that the phenomenon of extreme scattering is then restricted to clumps that lie well inside the star's Hill surface, so that very small impact parameters are required in order for extreme scattering to be observed. At present we lack a reliable theoretical basis for predicting plasma densities, and hence scattering characteristics, from the properties of the central star. In this paper our focus has been on O, B and A-type stars, implying that spectral types F, and later, are too cool to promote extreme scattering. That is a reasonable assumption, given the rapid change in ionizing luminosity near that boundary, but at present it is just an hypothesis.

Continuing to even cooler stars, we can presumably reach a regime where there are no molecular clumps that have plasma dense enough to cause ESEs and IDVs. However, molecular gas has a refractive index of its own, and self-gravitating molecular clouds can constitute strong lenses even without a plasma skin \citep{draine1998}. At frequencies well below the strong, far UV absorptions of H$_2$ and He, this type of lensing is almost achromatic. Thus, the molecular clump populations around cool stars might reveal themselves by radio-wave scintillations that are frequency-independent. In connection with that possibility, we note that weakly-chromatic scintillations have been reported in the blazar PKS1413+135 \citep{vedantham2017}, for which the line-of-sight passes close to Arcturus ($\alpha$ Boo, K0III, $z=11.3\,{\rm pc}$, $\rho=1.1\,{\rm pc}$).

It is unclear what role, if any, the star HD112934 plays in the observed scintillations of PKS1257-326. We account for the IDV of PKS1257-326 by the presence of Alhakim in the foreground: HD112934 is not needed. Nor is it a good match to the scintillation characteristics of PKS1257-326: $\beta^*=10.7^\circ$, approximately $64^\circ$ from $\beta$; $v_\perp^*=30.9\pm1.5\,{\rm km\,s^{-1}}$, yielding $p_\perp=0.32$; and $z^*=55\pm2\,{\rm pc}$, which is much larger than the $13.5\,{\rm pc}$ expected for the scattering material. These attributes all suggest that HD112934 is a random interloper, unrelated to the behaviour of PKS1257-326. 

What is surprising, though, is that HD112934 is so close to the line-of-sight: its impact parameter is $\rho^*=0.160\pm0.006\,{\rm pc}$. That is so small that the probability of this star being there by chance, computed using equation 1, is only about twice that calculated for Alhakim, even though $\Delta\beta$, $\Delta z$ and $p_\perp$ are all much larger. However HD112934 fits into the picture, it does nothing to weaken our conclusion that hot, foreground stars are central to the IDV phenomenon.

Having established the association between foreground stars and two particular IDV sources, it is sensible to examine the possible role of foreground stars more broadly, in extreme scattering phenomena. A thorough treatment is beyond the scope of this paper, but some brief commentary is appropriate. The two IDV sources we focused on were chosen because they offer precise localisations of the scattering plasma. No other known IDVs approach the level of precision of these two. Consequently, although it might be possible to identify plausible stellar hosts for the foreground plasma, we do not expect that associations could be made so confidently for any of the other known IDVs.

A good example is PKS0405-385, whose IDV has been intensively studied \citep{kedziora1997, kedziora2006}. Like J1819+3845 and PKS1257-326, its variations are very rapid, so we may reasonably expect the responsible plasma to be very local. Indeed, as PKS0405-385 is almost an order of magnitude brighter than either J1819+3845 or PKS1257-326, its angular size should be expected to be $\sim3\times$ larger, and if $v_\perp$ is not too different from the other two IDV then with a similar timescale we expect similar $a_\perp$, and hence $z_s$ should be smaller by a factor $\sim3$. That is so close that Sirius ($\alpha$ CMa, A1V, $z=2.1\,{\rm pc}$, $\rho=1.8\,{\rm pc}$) is the only plausible hot stellar counterpart. To determine whether Sirius is indeed the host, we need measurements of $\beta$, $v_\perp$ and $a_\perp$, for the plasma causing the scintillations. Unfortunately, although an evolution of the scintillation timescale of PKS0405-385 was noted by \citet{kedziora2006}, it was based on two short periods of variability, and no annual cycle has been demonstrated for this source. We therefore cannot confirm Sirius as the host star.

The same is true for the other extreme scattering phenomena, and for the same reason: poor localisation of the plasma. As an example: one of the best studied manifestations of parabolic arcs is in the pulsar B0834+06 \citep{stinebring2001,hill2005,brisken2010}. In that case the orientation, on the sky, of the plasma anisotropy is determined to similar precision as for our two IDV sources ($\beta$, in our Table 1). But the plasma velocity and line-of-sight distance both have large uncertainties --- the latter being several hundred parsecs \citep{brisken2010}. 

We expect similar difficulties in the case of IDV that is less extreme than that exhibited by PKS1257-326 and J1819+3845. Slower variations, with lower amplitude, are expected to arise when the scattering plasma is more distant, as a result of smoothing associated with the source size, and such variability should be relatively common. That is qualitatively consistent with the many cases of IDV reported by \citet{lovell2008}. Identifying the stellar host of the scattering plasma will be more difficult in those cases, because the larger range in line-of-sight distance admits more potential counterparts. However, if firm identifications prove elusive, a statistical model may be worthwhile --- using the known locations and velocities of hot stars in the solar neighbourhood to predict the chances of observing IDV in each radio source.

In the introduction to this paper we noted that IDVs and ESEs are likely to be related, but we did not say how. Now, armed with a picture of the plasma concentrations responsible for IDV, we can suggest a possible relationship. Consider a cometary knot, such as in the Helix. As the width of the knot is comparable to the extent of the plasma responsible for IDV in J1819+3845 \citep{debruynmacquart2015}, we suppose that IDV arises when the line-of-sight pierces a typical point, suggesting that microstructure is present throughout the plasma skin. The ESE phenomenon, however, does not seem to be caused by microstructure; instead it requires a column-density curvature that is both strong and consistent in its sign over a broad spatial extent \citep{tuntsov2016}. For a thin skin of plasma, the smooth component of the  column-density -- i.e. its macrostructure -- has a curvature that is strongest for lines-of-sight that are just inside the boundary of the structure, so that is where we might plausibly expect ESEs to arise.

Various types of studies would be useful in testing our ideas. Atomic emission lines, such as H$\alpha$, are generically expected to arise from extreme scattering plasma, so their discriminating power is modest. Imaging in molecular emission lines, on the other hand, can directly address some of the key issues raised in this paper. Taking the Helix as our paradigm motivates observations in the H$_2$ rovibrational lines, for example, and the pure rotational lines of CO. We note, however, the factor $\sim10$ difference in photospheric temperatures between the upper main sequence and the central star of the Helix. Therefore, if it is ionizing photons that are  responsible for creating the cometary knots, we expect that their optical/near-IR emission will appear very faint in comparison with any main-sequence host.

\section{Conclusions}
Position and velocity constraints on the plasma responsible for IDV in J1819+3845 and PKS1257-326 place the material close to Vega and Alhakim, respectively --- two nearby A-stars. The probability that this coincidence occurred by chance is very low and we conclude that the relationship is causal: the scattering material must be associated with these stars. The size, location, orientation, and numbers of plasma concentrations all point to circumstellar environments that are fundamentally similar to the Helix Nebula, where it is known that the plasma is associated with dense molecular clumps. Therefore we conclude that similar populations of molecular clumps are present around hot stars in the solar neighbourhood. Bearing in mind the selection bias associated with IDV studies -- i.e. rapid scintillation of radio quasars requires nearby, high-density plasma -- we anticipate that circumstellar molecular clumps are likely to be ubiquitous across all stellar spectral types, and throughout the Galaxy. If so, three independent estimates indicate that those clumps make a major contribution to the mass of the Galaxy.

 \acknowledgments
Hats off to Ger de Bruyn and Jane Dennett-Thorpe for spotting, very early on, the possibility of a connection between Vega and J1819+3845. The referee's thoughtful comments helped to sharpen our presentation.


\begin{thebibliography}{}
\bibitem[Antonov and Latyshev(1972)]{antonovlatyshev1972}Antonov, V.A., Latyshev, I.N., 1972, IAUS, 45, 341
\bibitem[Armstrong, Rickett and Spangler(1995)]{armstrong1995}Armstrong, J.W., Rickett, B.J., Spangler, S.R., 1995, ApJ, 443, 209
\bibitem[Bannister et al(2016)]{bannister2016}Bannister, K.W., Stevens, J., Tuntsov, A.V., et al. 2016, Science, 351, 354
\bibitem[Bannister et al(2017)]{bannister2017}Bannister, K.W., Bignall, H.E., Johnston, S., et al. 2017, ATel 10024
\bibitem[Bignall et al(2003)]{bignall2003}Bignall,~H.E., Jauncey,~D.L., Lovell,~J.E.J., et al. 2003 ApJ 585, 653
\bibitem[Bignall et al(2006)]{bignall2006}Bignall,~H.E., Macquart,~J.P., Jauncey,~D.L., et al.  2006 ApJ 652, 1050
\bibitem[Bovy(2017)]{bovy2017}Bovy, J., 2017, MNRAS, 468, L63
\bibitem[Brisken et al(2010)]{brisken2010}Brisken, W.F., Macquart, J.-P., Gao, J.J., et al. 2010, ApJ, 708, 232
\bibitem[Cordes and Wolszczan(1986)]{cordeswolszczan1986}Cordes, J.M., Wolszczan, A., 1986, ApJ, 307, L27
\bibitem[Cordes et al(2006)]{cordes2006}Cordes,~J.M., Rickett,~B.J., Stinebring,~D.R., Coles,~W.A. 2006 ApJ 637, 346
\bibitem[Cr\'ez\'e et al(1998)]{creze1998}Cr\'ez\'e, M., Chereul, E., Bienaym\'e, O., Pichon, C., 1998, A\&A, 329, 920
\bibitem[de~Bruyn and Macquart(2015)]{debruynmacquart2015}de~Bruyn, A.G., Macquart, J.-P., 2015, A\&A, 574, 125
\bibitem[Dennett-Thorpe and de~Bruyn(2000)]{dennettthorpe2000}Dennett-Thorpe,~J., de~Bruyn, A.G., 2000, ApJL, 529, L65
\bibitem[Dennett-Thorpe and de~Bruyn(2002)]{dennettthorpe2002}Dennett-Thorpe,~J., de~Bruyn, A.G., 2002, Nature, 415, 57
\bibitem[Dennett-Thorpe and de~Bruyn(2003)]{dennettthorpe2003}Dennett-Thorpe,~J., de~Bruyn, A.G., 2003, A\&A 404, 113
\bibitem[Draine(1998)]{draine1998}Draine, B.T., 1998, ApJL, 509, L41
\bibitem[Fiedler et al(1987)]{fiedler1987}Fiedler, R. L., Dennison, B., Johnston, K.J., Hewish, A., 1987, Nature, 326, 675
\bibitem[Fiedler et al(1994)]{fiedler1994}Fiedler, R., Dennison, B., Johnston, K.J., Waltman, E.B., Simon, R.S., 1994, ApJ, 430, 581
\bibitem[Garc\'ia-Segura et al(2006)]{garciasegura2006}Garc\'ia-Segura, G., L\'opez, J.A., Steffen, W., Meaburn, J., Manchado, A., 2006, ApJ, 646, L61
\bibitem[Goldreich and Sridhar(1995)]{goldreichsridhar1995}Goldreich, P., Sridhar, S., 1995, ApJ, 438, 763
\bibitem[Heisler and Tremaine(1986)]{heislertremaine1986}Heisler, J., Tremaine, S., 1986, Icarus, 65, 13
\bibitem[Hill et al(2005)]{hill2005}Hill, A., Stinebring, D.R., Asplund, C.T., et al. 2005, ApJ, 619, L171
\bibitem[Huggins et al(1992)]{huggins1992}Huggins, P.J., Bachiller, R., Cox, P., Forveille, T., 1992, ApJ, 401, L43
\bibitem[Huggins et al(2002)]{huggins2002}Huggins, P.J., Forveille, T., Bachiller, R., et al. 2002, ApJ, 573, L55
\bibitem[Kedziora-Chudczer et al(1997)]{kedziora1997}Kedziora-Chudczer~L., Jauncey~D.L., Wieringa~M.H., et al 1997, ApJL, 490, L9
\bibitem[Kedziora-Chudczer(2006)]{kedziora2006}Kedziora-Chudczer, L., 2006, MNRAS, 369, 449
\bibitem[Lanz and Catala(1992)]{lanzcatala1992}Lanz, T., Catala, C., 1992, A\&A, 257, 663
\bibitem[Linsky, Rickett and Redfield(2008)]{linsky2008}Linsky, J.L, Rickett, B.J., Redfield, S., 2008, ApJ, 675, 413
\bibitem[Lovell et al(2003)]{lovell2003}Lovell, J.E.J., Jauncey, D.L., Bignall,~H.E., et al. 2003, AJ, 126, 1699
\bibitem[Lovell et al(2008)]{lovell2008}Lovell, J.E.J., Rickett, B.J., Macquart, J.-P., et al. 2008, ApJ, 689, 108
\bibitem[Matsuura et al(2009)]{matsuura2009}Matsuura, M., Speck, A.K., McHunu, B.M., et al. 2009, ApJ, 700, 1067
\bibitem[Meaburn et al(1992)]{meaburn1992}Meaburn, J., Walsh, J.R., Clegg, R.E.S., et al. 1992, MNRAS, 255, 177
\bibitem[Meaburn et al(1998)]{meaburn1998}Meaburn, J., Clayton, C.A., Bryce, M., et al. 1998, MNRAS, 294, 201
\bibitem[Meixner et al(2005)]{meixner2005}Meixner, M., McCullogh, P., Hartman, J., Son, M., Speck, A., 2005, AJ, 130, 1784
\bibitem[O'Dell and Handron(1996)]{odellhandron1996}O'Dell, C.R., Handron, K.D.,1996, AJ, 111, 1630
\bibitem[O'Dell et al(2002)]{odell2002}O'Dell, C.R., Balick, B., Hajian, A.R., Henney, W.J., Burkert, A.R., 2002, AJ, 123, 3329
\bibitem[Parker(1958)]{parker1958}Parker, E.N., 1958, ApJ, 128, 664
\bibitem[Pen and Levin(2014)]{penlevin2014}Pen, U.-L., Levin, Y., 2015, MNRAS, 442, 3338
\bibitem[Perryman et al(1997)]{perryman1997}Perryman, M.A.C., Lindegren, L., Kovalevsky, J., et al. 1997, A\&A, 323, L49
\bibitem[Quanz et al(2011)]{quanz2011}Quanz, S.P., Kenworthy, M.A., Meyer, M.R., Girard, J.H.V., Kasper, M., 2011, ApJ, 736, L32
\bibitem[Rickett(1990)]{rickett1990}Rickett, B.J., 1990 ARAA 28, 561
\bibitem[Rickett, Lyne and Gupta(1997)]{rickett1997}Rickett, B.J., Lyne, A.G., Gupta, Y., 1997, MNRAS, 287, 739
\bibitem[Rickett, Kedziora-Chudczer and Jauncey(2002)]{rickett2002}Rickett,~B.J., Kedziora-Chudczer,~L., Jauncey,~D.L. 2002 ApJ 581, 103
\bibitem[Stinebring et al(2001)]{stinebring2001}Stinebring, D.R., McLaughlin, M.A., Cordes, J.M., et al. 2001, ApJ, 549, L97
\bibitem[Tuntsov, Bignall and Walker(2013)]{tuntsov2013}Tuntsov, A.V., Bignall, H.E., Walker, M.A., 2013, MNRAS, 429, 2562
\bibitem[Tuntsov et al(2016)]{tuntsov2016}Tuntsov, A.V., Walker, M.A., Koopmans, L.V.E., et al. 2016, ApJ, 817, 176
\bibitem[Tuntsov et al(2017)]{tuntsov2017}Tuntsov, A.V., Stevens, J., Bannister, K.W., et al. 2017, MNRAS (submitted)
\bibitem[Vedantham et al(2017)]{vedantham2017}Vedantham, H.K., Readhead, A.C.S., Hovatta, T., et al. 2017, ApJ (submitted), arXiv:1702.05519 
\bibitem[Walker and Wardle(1998)]{walkerwardle1998}Walker, M.A., Wardle, M.J., 1998 ApJ, 498, L125
\bibitem[Walker et al(2004)]{walker2004}Walker,~M.A., Melrose,~D.B., Stinebring,~D.R., Zhang,~C.M., 2004 MNRAS 354, 43
\bibitem[Walker(2007)]{walker2007}Walker, M.A., 2007, ``SINS - Small Ionized and Neutral Structures in the Diffuse ISM'' Eds.: M. Haverkorn, W.M. Goss (San Francisco: ASP) ASP Conf. Ser., 365, 299
\bibitem[Walker, de~Bruyn and Bignall(2009)]{walker2009}Walker, M.A., de Bruyn, A.G., Bignall, H.E., 2009, MNRAS, 397, 447

\end{thebibliography}
\end{document}